\documentclass [twocolumn, aps, prl, superscriptaddress, showpacs] {revtex4}
\usepackage{graphicx,latexsym}
\begin{document}
\draft
\title {Giant frictional drag in strongly interacting bilayers near filling factor one}

\author {E. Tutuc}
\affiliation {Department of Electrical Engineering, Princeton
University, Princeton, NJ 08544}
\affiliation {Microelectronics
Research Center, University of Texas at Austin, Austin, TX 78758}
\author {R. Pillarisetty}
\affiliation{Department of Electrical Engineering, Princeton
University, Princeton, NJ 08544}
\author {M. Shayegan}
\affiliation{Department of Electrical Engineering, Princeton
University, Princeton, NJ 08544}
\date{\today}
\begin{abstract}
We study the frictional drag in high mobility, strongly
interacting GaAs bilayer hole systems in the vicinity of the
filling factor $\nu=1$ quantum Hall state (QHS), at the same
fillings where the bilayer resistivity displays a reentrant
insulating phase. Our measurements reveal a very large
longitudinal drag resistivity ($\rho^{D}_{xx}$) in this regime,
exceeding 15 k$\Omega/\Box$ at filling factor $\nu=1.15$.
$\rho^{D}_{xx}$ shows a weak temperature dependence and appears to
saturate at a finite, large value at the lowest temperatures. Our
observations are consistent with theoretical models positing a
phase separation, e.g. puddles of $\nu=1$ QHS embedded in a
different state, when the system makes a transition from the
coherent $\nu=1$ QHS to the weakly coupled $\nu=2$ QHS.
\end{abstract}
\pacs{73.43.-f, 71.35.-y, 73.22.Gk} \maketitle

Closely spaced bilayer carrier systems have been the test ground
for a multitude of novel electronic states with no counterpart in
the single-layer case. The most important are quantum Hall states
(QHSs) possessing inter-layer coherence \cite{wenzee} at total
filling factor $\nu=1/2$ (layer filling $\nu_{layer}=1/4$)
\cite{suen, eisen} and $\nu=1$ ($\nu_{layer}=1/2$)
\cite{murphy,lay}. These QHSs are stabilized when the interaction
between carriers in the same layer is comparable to that of
carriers residing in opposite layers. The $\nu=1$ QHS has been
shown to exhibit enhanced inter-layer tunnelling \cite{spielman}
reminiscent of a Josephson junction, as well as a peculiar,
charge-neutral superfluid in counterflow transport \cite{kellogg,
tutuc}. In a simple picture the $\nu=1$ QHS can be regarded as a
condensate of excitons \cite{yang}, where carriers and vacancies
pair-up in the opposite, half-filled layer forming excitons, which
condense at lowest temperatures.

An equally interesting ground state also explored in conjunction
with the emergence of high quality, interacting bilayer systems is
the Wigner crystal (WC) \cite{shayegan}. Experimentally, transport
measurements in electron and hole bilayers show a suppression of
QHSs beyond a given filling factor, namely $\nu=1/2$ in
interacting electron bilayers \cite{hari} and $\nu=1$ in
interacting hole bilayers \cite{tutuc_2003}. This observation is
similar to the suppression of fully developed QHSs in single
layers beyond $\nu=1/5$ for electrons, and $\nu=1/3$ for dilute
holes which has been interpreted as a signature of the WC being
stabilized at sufficiently low fillings \cite{shayegan}.
Furthermore, the quenching of QHSs at sufficiently low fillings is
accompanied by the existence of a reentrant insulating phase (RIP)
around the lowest filling QHS, suggesting an onset of the WC
state. In order to gain further insight into the physics of the
RIP, here we study the frictional drag in interacting GaAs hole
bilayers in the vicinity of the phase coherent $\nu=1$ QHS, in the
same filling factor range where the bilayer resistivity exhibits a
RIP. Our results show an anomalously, record large longitudinal
drag resistivity ($\rho^{D}_{xx}$) on the flanks of $\nu=1$,
larger than 15 k$\Omega/\Box$. Equally anomalous is the relatively
weak temperature dependence of $\rho^D_{xx}$; it follows a power
law $\rho^{D}_{xx}\sim T^{\alpha}$, with $\alpha<1$, and saturates
at a finite value at the lowest temperatures.

Our sample is a Si-modulation-doped GaAs double-layer hole system
grown on GaAs (311)A substrate. It consists of two GaAs quantum
wells which have a width of 150$\AA$ each and are separated by a
75$\AA$ wide AlAs barrier. The sample is patterned in a Hall bar
geometry of $100\mu$m width, aligned along the [01$\bar{1}$]
crystal direction \cite{anisotropy}. Diffused InZn ohmic contacts
are placed at the end of each lead. We use a combination of front
and back gates \cite{eisenstein-apl} to selectively deplete one of
the layers near each contact. As grown, the densities in the two
layers were $p_{T}=2.6\times10^{10}$ cm$^{-2}$ and
$p_{B}=3.2\times10^{10}$ cm$^{-2}$ for the top and bottom layers,
respectively. The mobility along [01$\bar{1}$] at these densities
is approximately 200,000 cm$^2/$Vs. Metallic top and bottom gates
are added on the active area to control the densities in the
layers. The measurements are performed down to a temperature of
$T=30$mK, and using standard low-current (0.5nA-1nA),
low-frequency lock-in techniques.

Two types of measurement configurations are used in our study. In
one (bilayer) configuration, current is passed through both top
and bottom layers and the ohmic contacts connect both layers
simultaneously. The voltage drops along and across the Hall bar,
divided by the total bilayer current, represent the longitudinal
($\rho^{B}_{xx}$) and Hall ($\rho^{B}_{xy}$) bilayer
resistivities. In a second (drag) configuration, current is passed
in one (drive) layer only, by using the selective depletion
technique around the ohmic contacts such that they connect to a
single layer only \cite{eisenstein-apl}. The voltage drops
measured in the opposite (drag) layer, divided by the drive
current, represent the longitudinal ($\rho^{D}_{xx}$) and Hall
($\rho^{D}_{xy}$) drag resistivities. The drag resistivity
provides a measure of the electron-electron scattering rate
between the carriers in the drive layer and those in the drag
layer. For the data presented here we adopt the following sign
convention: the longitudinal (Hall) drag resistivity is defined as
positive when the voltage drop along (across) the drag layer is
{\it opposite} to the voltage drop along (across) the drive layer.
We performed the usual consistency checks associated with drag
measurements \cite{gramilla}. Owing to the proximity of the two
layers in our sample, there is a small but finite inter-layer
leakage current.  This leakage translates into an uncertainty in
frictional drag measurements, which does not exceed $\pm6\%$ in
our study.

In the main panel of Fig. 1 we show $\rho^{B}_{xx}$ and
$\rho^{D}_{xx}$ vs the applied perpendicular magnetic field ($B$),
both measured at $T=30$mK. The total bilayer density is
$p_{tot}=5.5\times10^{10}$ cm$^{-2}$, equally distributed between
the two layers (balanced). The data show a fully developed QHS at
$\nu=1$, stabilized here solely by interlayer coherence
\cite{tunneling}. In a simple picture the emergence of a QHS at
$\nu=1$ can be understood by considering the pairing of carriers
and vacancies in the opposite layers. At total filling factor one
each layer has the lowest Landau level half full, i.e. has an
equal number of carriers and vacancies. Owing to the close
proximity of the two layers and the ensuing inter-layer
interaction, it is energetically favorable to form carrier-vacancy
pairs in the opposite layers, which condense at the lowest
temperature. A spectacular signature of this phenomenon is the
emergence of a neutral superfluid, experimentally observed when
equal and opposite currents are passed in the two layers
\cite{kellogg,tutuc}. The ratio between the interaction energy of
carriers in different layers and in the same layer is commonly
quantified by $d/l_{B}$, where $d$ is the interlayer distance and
$l_{B}=\sqrt{\hbar/eB}$ is the magnetic length at $\nu=1$. For the
case examined in Fig. 1 this ratio is 1.33.

\begin{figure}
\centering
\includegraphics[scale=0.63]{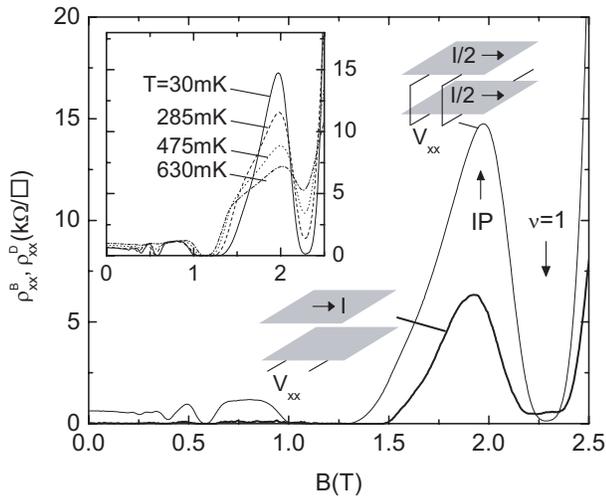}
\caption {\small{ Bilayer and longitudinal drag resistivities
($\rho^{B}_{xx}$ and $\rho^{D}_{xx}$) measured at $T=30$mK for a
balanced bilayer with $p_{tot}=5.5\times10^{10}$ cm$^{-2}$. Note
that both traces are plotted on the same scale. The inset shows
the temperature dependence of $\rho^{B}_{xx}$ data.}}
\end{figure}

The inset of Fig. 1 shows $\rho^{B}_{xx}$ measured at different
temperatures, for the same layer densities as in the main panel.
These data show that as the temperature is reduced a RIP develops
on the flanks of the $\nu=1$ QHS. Most interestingly, the data of
Fig. 1 show a very large longitudinal drag on the left flank of
$\nu=1$, in the same filling factor range where $\rho^{B}_{xx}$
exhibits a RIP. In contrast to typical drag measurements where the
drag resistivity is one to three orders of magnitude smaller than
the single layer resistance \cite{gramilla}, Fig. 1 data reveal
that $\rho^{D}_{xx}$ and $\rho^{B}_{xx}$ are of the {\it same}
order or magnitude, which testifies to the strong interlayer
coupling at these filling factors. Clearly frictional drag
constitutes a substantial component of the longitudinal
resistivity here, in contrast to frictional drag at $B=0$T where
drag is a very small perturbation.

\begin{figure}
\centering
\includegraphics[scale=0.75]{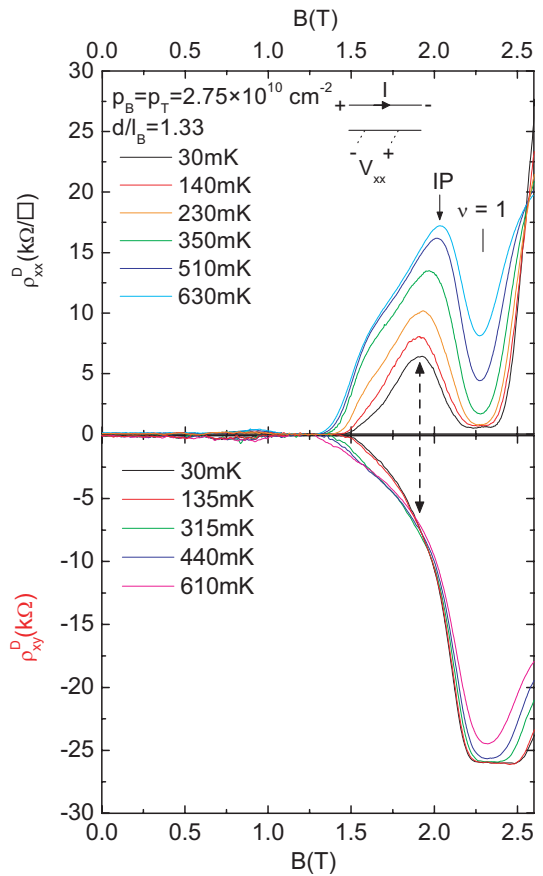}
\caption {\small{(color online) $\rho^{D}_{xx}$ (top panel) and
$\rho^{D}_{xy}$ (bottom panel) vs $B$ measured at different
temperatures. The total bilayer density is
$p_{tot}=5.5\times10^{10}$ cm$^{-2}$, equally distributed in the
two layers. The data reveal a very large $\rho^{D}_{xx}$ on the
left flank of $\nu=1$, concomitant with the observed RIP.}}
\end{figure}

In Fig. 2 we show $\rho^{D}_{xx}$ (top panel) and $\rho^{D}_{xy}$
(bottom panel) vs $B$, measured at different temperatures ranging
from 30mK to 630mK, and at the same layer densities as the data of
Fig. 1. At the lowest temperatures the data of Fig. 2 show a
nearly vanishing $\rho^{D}_{xx}$ at $\nu=1$ and $\rho^{D}_{xy}$
quantized at $h/e^{2}=25.88$ k$\Omega$ \cite{kellogg2002}. The
temperature dependence of the Hall drag measured at and around
$\nu=1$ is relatively weak: as $T$ is increased $\rho^{D}_{xy}$
remains close to the quantized value for $T$ as high as 500mK. The
weak temperature dependence of $\rho^D_{xy}$ at $\nu=1$ is
consistent with previous results in GaAs hole bilayers, which show
a vanishing counterflow Hall resistivity for temperatures below
500mK \cite{tutuc} and indicates a strong pairing of the carriers
and vacancies in opposite layers. Figure 2 data (top panel)
substantiates our observation of an anomalously large drag in the
vicinity of $\nu=1$ QHS. As the temperature is increased
$\rho^{D}_{xx}$ increases, reaching a record 17k$\Omega/\Box$ at
$T=630$mK at $\nu=1.10$. Equally noteworthy is that the onset of
the anomalously large $\rho^{D}_{xx}$ {\it coincides} with the
onset of the non-zero $\rho^{D}_{xy}$, indicating that the
particle-vacancy pairing which stabilizes the $\nu=1$ QHS is also
responsible for the observed anomalously large $\rho^{D}_{xx}$.

\begin{figure}
\centering
\includegraphics[scale=0.33]{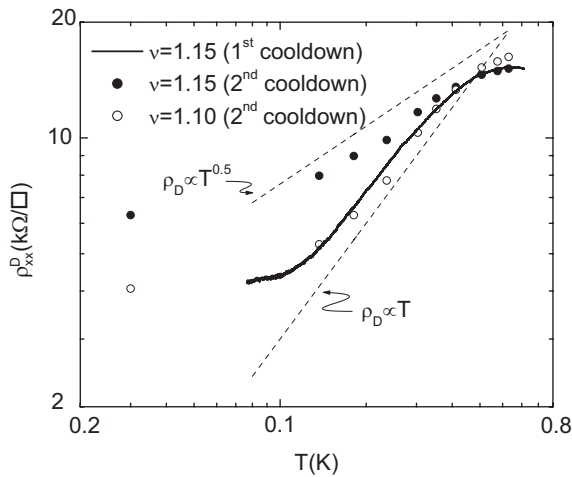}
\caption {\small{Temperature dependance of $\rho^{D}_{xx}$
measured at $\nu=1.10$ and $\nu=1.15$ in two different cooldowns.
The different temperature dependances in separate cooldowns
suggest that sample disorder affects the measured
$\rho^{D}_{xx}$.}}
\end{figure}

Next we present the temperature dependence of the anomalously
large longitudinal drag observed near $\nu=1$. Figure 3 shows the
$\rho^{D}_{xx}$ vs $T$ data from 30mK to 630mK. An increase in
interlayer current prevents an accurate measurement of the
frictional drag near $\nu=1$ above $T=700$mK. The $\rho^{D}_{xx}$
data were measured in two different cooldowns at $\nu=1.10$ and
$\nu=1.15$, namely fillings where the RIP reaches the maximum
resistance. Note that $\rho^{D}_{xx}$ maximum shifts slightly from
$\nu=1.15$ at the lowest $T$ to $\nu=1.10$ at the highest $T$, as
apparent from Fig. 2(top panel) data. Several features of Fig. 3
data are noteworthy. First, $\rho^{D}_{xx}$ exhibit a weak,
slightly sublinear temperature dependence in the range
$T=100-500$mK, which contrasts the more common
$\rho^{D}_{xx}\propto T^{2}$ characteristic of the Coulomb drag in
two-dimensional electron systems \cite{gramilla} or
$\rho^{D}_{xx}\propto T^{4/3}$ observed in drag measurements
between composite fermions \cite{lilly}. Second, $\rho^{D}_{xx}$
appears to saturate at a constant, {\it finite} value below
$T=100$mK \cite{heating}. Third, the large $\rho^{D}_{xx}$ near
$\nu=1$ shows a cooldown dependence, suggesting that sample
disorder affects the measured $\rho^{D}_{xx}$.

Before discussing our observation of enhanced frictional drag in
the vicinity of $\nu=1$ within existing theoretical models, we
summarize the salient features of the experimental data. First,
the longitudinal drag is greatly enhanced in the vicinity of the
bilayer $\nu=1$ QHS, exceeds 15 k$\Omega/\Box$, and becomes
comparable to the single-layer longitudinal resistivity. Second,
the observed giant longitudinal drag emerges concomitantly with
the large Hall drag near $\nu=1$, indicating that particle-vacancy
pairing is present. Third, the giant longitudinal drag has a weak,
sub-linear temperature dependence, and appears to saturate at a
finite, and large value ($\simeq$5k$\Omega/\Box$) at the lowest
temperatures. Finally, the frictional drag exhibits a cooldown
dependence, which suggests that disorder affects the measured
$\rho^{D}_{xx}$ value. These features contrast the frictional drag
between two two-dimensional carrier systems, which typically has a
small (~1-100$\Omega$) magnitude, and a $\rho^{D}_{xx}\propto
T^{2}$ temperature dependence \cite{gramilla}. And while our
measurements are performed in the quantum Hall regime where an
agreement with Fermi liquid theory \cite{price,zhang} should not
be expected, these highlighted differences are nonetheless stark.

\begin{figure}
\centering
\includegraphics[scale=0.42]{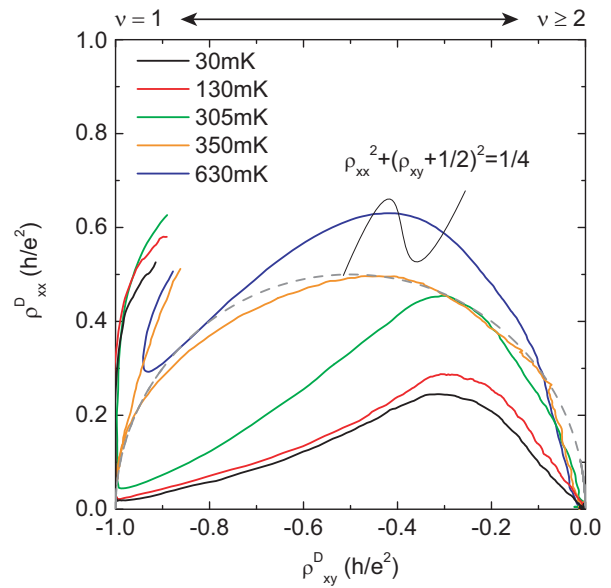}
\caption {\small{(color online) $\rho^{D}_{xx}$ vs $\rho^{D}_{xy}$
in units of $h/e^2$ for different temperatures. The end points
represent the weakly ($\nu\geq0$, $\rho^{D}_{xy}=0$) and strongly
($\nu=1$, $\rho^{D}_{xy}=-h/e^2$) coupled bilayer regimes. The
dashed line represents the semicircle relation
$(\rho^{D}_{xy}+1/2)^2+{\rho^{D}_{xx}}^2=1/4$ expected according
to Ref. 22.}}
\end{figure}

Our data can qualitatively be explained by theoretical models
which invoke the coexistence of two phases as the system makes a
transition, driven by filling factor in our case, from the $\nu=1$
QHS to the weakly coupled $\nu=2$ QHS. The $\nu=2$ QHS consists of
a pair of $\nu_{layer}=1$ QHSs, one in each of the two layers.
Stern and Halperin \cite{stern} examined theoretically the
transition between the strongly coupled $\nu=1$ QHS and two weakly
coupled layers, each at $\nu_{layer}=1/2$. By postulating that in
the transition regime the system is composed of puddles of $\nu=1$
QHS phase, and assuming the conductivity tensors in the two
regimes, namely a strongly coupled $\nu=1$ QHS on one hand and two
weakly coupled layers at $\nu=1/2$ each on the other, they derive
an expression for the longitudinal and Hall drag as a function of
the fraction of the $\nu=1$ QHS across the transition. Their model
predicts a large longitudinal drag, as high as $h/2e^2$ in the
transition regime, concomitantly with a non-zero Hall drag. Their
results can analytically be approximated by a simple semi-circle
relation for the drag resistivity tensor,
\begin{equation}
\centering (\rho^{D}_{xy}+1/2)^2+(\rho^{D}_{xx})^2=1/4
\end{equation}
with the resistivity expressed in units of $h/e^2$. Kellogg {\it
et al.} \cite{kellogg03} have experimentally probed this
transition by varying the total bilayer density, which in turn
changes $d/l_{B}$. They observe an enhanced longitudinal drag in
the transition region, in qualitative agreement with the
theoretical model \cite{stern}.

In order to quantitatively compare our experimental results with
the model of Ref. 22, in Fig. 4 we show $\rho^{D}_{xx}$ vs
$\rho^{D}_{xy}$ at different temperatures along with the
semi-circle law of Eq. (1). The end points of the semi-circle,
namely $\rho^{D}_{xx}=\rho^{D}_{xy}=0$ and $\rho^{D}_{xx}=0$,
$\rho^{D}_{xy}=-1$, represent the weakly and strongly coupled
bilayer regimes at $\nu\geq2$ and $\nu=1$, respectively. As the
system makes the transition from weakly to strongly coupled,
$\rho^{D}_{xx}$ and $\rho^{D}_{xy}$ depart from zero
simultaneously, with $\rho^{D}_{xx}$ reaching a temperature
dependent maximum. $\rho^{D}_{xx}$ is close to $h/2e^2=12.9$
k$\Omega/\Box$ predicted by Eq. (1). At intermediate temperatures
$T\approx300$mK the $\rho^{D}_{xx}$ vs $\rho^{D}_{xy}$ data are in
very good quantitative agreement with the semi-circle law of Eq.
(1), but depart from it at the lowest temperatures. We note
however that Eq. (1) is expected to hold quantitatively if the
drag resistivity is large compared to the symmetric (parallel
flow) bilayer resitivity at all fillings, and also neglects the
bilayer and drag resistivities in the weakly coupled regime. In
light of these approximations, the agreement with the simple
semi-circle law is satisfactory.

A separate model, also invoking the co-existence of two phases,
that may explain the giant frictional drag data has been proposed
by Spivak and Kivelson \cite{spivak}. The model of Ref. 24
considers the frictional drag between a passive layer and a
low-density two-dimensional system where the ground state consists
of bubbles of Wigner crystal (WC) embedded in a Fermi liquid. Each
WC bubble in the active layer casts an image in the passive layer,
which can be pictured as a hard wall potential being dragged in
the passive layer. This in turn results in a significant
scattering for the electrons in the passive layer, hence an
anomalously large frictional drag. We speculate that one plausible
scenario for the large drag in the vicinity of $\nu=1$ in our
sample is a "micro-emulsion" of the $\nu=1$ QHS co-existing with a
WC state.

In summary we report the observation of giant frictional drag in
the vicinity of the strongly coupled bilayer $\nu=1$ QHS. The
giant longitudinal drag emerges concomitantly with a non-zero Hall
drag, indicating the particle-vacancy pairing in this regime. Our
observations are consistent with theoretical models
\cite{stern,spivak} which invoke the co-existence of two distinct
phases as the system makes a transition from the $\nu=1$ bilayer
QHS, e.g. puddles of $\nu=1$ QHS embedded in a weakly coupled bulk
state or in a Wigner crystal state.

We thank Ady Stern for discussions, and DOE, NSF, and SWAN-NRI
center for support.


\begin{thebibliography}{10}
\bibitem{heating}
While took precautions to use sufficiently low currents to avoid
carrier Joule heating, we cannot rule out a lack of full carrier
thermalization at the lowest temperatures.
\end{thebibliography}
\end{document}